\documentclass[aps,pre,amssymb,twocolumn,superscriptaddress]{revtex4-1}
\usepackage{graphicx}
\usepackage{amsmath}

\begin{document}
\title{An interactive city choice model and its application for measuring the intercity interaction}
\author{Xiang-Yu Jia}
\affiliation{Key Laboratory of Transport Industry of Big Data Application Technologies for Comprehensive Transport, Ministry of Transport, Beijing Jiaotong University, Beijing {\rm 100044}, China}

\author{Er-Jian Liu}
\email{erjianliu@bjtu.edu.cn}
\affiliation{Institute of Transportation System Science and Engineering, Beijing Jiaotong University, Beijing {\rm 100044}, China}

\author{Chun-Yan Chen}
\affiliation{Traffic Control Technology Co.,Ltd, Beijing {\rm 100875}, China}

\author{Xiao-Yong Yan}
\email{yanxy@bjtu.edu.cn}
\affiliation{Institute of Transportation System Science and Engineering, Beijing Jiaotong University, Beijing {\rm 100044}, China}

\begin{abstract}
Measuring the interaction between cities is an important research topic in many disciplines, such as sociology, geography, economics and transportation science. The traditional and most widely used spatial interaction model is the gravity model, but it requires the parameters to be artificially set. In this paper, we propose a parameter-free interactive city choice (ICC) model that measures intercity interaction from the perspective of individual choice behavior. The ICC model assumes that the probability of an individual choosing to interact with a city is proportional to the number of opportunities in the destination city and inversely proportional to the number of intervening opportunities between the origin city and the destination city, calculated using the travel time in the transportation network. The intercity interaction intensity can be obtained by calculating the product of this probability and the origin city's population. We apply the ICC model to measure the interaction intensity among 339 cities in China and analyze the impact of changes in the Chinese land transportation network from 2005 to 2018 on the intercity and city interaction intensity. Compared with the previously widely used spatial interaction models, the measurement results of the ICC model are more consistent with the actual situation.
\end{abstract}


\maketitle
	\section*{Introduction}
	The rapid development of cities worldwide and the acceleration of urbanization have led to more than half of the world's population living in cities\cite{RitchieRoser-482},and thus cities have become the main location for human activities in today's society\cite{Barthelemy-550}. The connections between cities through the transportation network promotes the flow of people, goods, information, money and skills among cities; such flow between cities is called intercity interaction\cite{Ord-552,Besag-549}. Understanding and predicting intercity interaction patterns has long been an important research topic in sociology, geography, economics, transportation science and many other disciplines\cite{Wilson-560,najem2018debye}. It also has great significance in the rational formulation of urban development strategies\cite{Batty-486,paldino2015urban}, the promotion of regional sustainable development\cite{GatelyHutyra-556}, communicable disease control\cite{HufnagelBrockmann-490,EubankGuclu-491,BalcanColizza-492} and other fields. As the intercity interaction intensity increases, cities are no longer regarded as isolated individuals but as interdependent urban systems\cite{Batten-553}. Therefore, understanding the intercity interaction and establishing a model that can accurately measure the interaction between cities are of great value for optimizing the spatial structure of urban agglomerations\cite{Barthelemy-554,LeNEChet-555}.
	
	The gravity model was the first model proposed to measure intercity interaction\cite{Stewart-557}. The model assumes that the intensity of interaction between two cities is proportional to the product of their sizes (e.g., population, GDP) and inversely proportional to a power law function of their distance. The gravity model is simple in form and is widely used to predict intercity interactions, such as intercity travel\cite{JungWang-499}, commuting trips\cite{ViboudBjORnstad-501,lenormand2016systematic}, population migration\cite{Tobler-487} and international trade\cite{Fagiolo-567}. However, this model is based on analogy with Newton's law of universal gravitation and does not involve individual spatial choice behavior\cite{Wilson-560,YanZhou-508}. Furthermore, the parameter of the gravity model's power law distance function is artificially defined. For example, some researchers set the parameter to 1\cite{GoldenbergLevy-564,backstrom2010find,rahman2009australia}, while others  set it to 2\cite{KringsCalabrese-534,Keum-577,AlamUddin-578}. Therefore,  it would be a valuable contribution to establish a parameter-free model to measure intercity interactions from the perspective of individual spatial choice behavior.
	
	Simini et al. took an important step forward in spatial interaction modeling by establishing a parameter-free model named the radiation model\cite{SiminiGonzALez-509} to predict commuting trips between counties in the U.S. This model assumes that the individual will consider the employment opportunities provided by the work location and the benefits that the opportunities may bring to him/her when choosing a work location. He/she will choose the work location nearest to his/her home that offers a benefit greater than the best offer available in his/her home county. 
	Some researchers improve the radiation model and propose various commuting prediction models, such as the radiation model with selection\cite{simini2013human} and the flow and jump model\cite{varga2018commuting}.
	Recently, many researchers have applied the radiation model or improved radiation models to measure intercity interaction intensity\cite{LiFeng-561,Tian-562,ZhengKuang-563}. However, the radiation model assumes that the individual will only choose the nearest location with a higher benefit than his/her home, which reflects a cautious tendency of individual choice behavior. It can predict commuting trips but is not suitable for predicting general travel\cite{LiuYan-566} because travelers may choose not only the closest location with a higher benefit than the origin but also other locations with higher benefits than the origin and intervening destinations . To solve this problem, Liu and Yan proposed another parameter-free model, named the opportunity priority selection (OPS) model\cite{LiuYan-519}, that adopts the perspective of individual destination choice behavior. The OPS model assumes that when the individual chooses a destination, he/she will choose a location with a higher benefit than the benefit of the origin and the benefits of the intervening opportunities\cite{Stouffer-498}. This reflects an exploratory tendency in individual choice behavior and can accurately predict human mobility within and between cities. Compared with the radiation model, the OPS model can better describe individual destination choice behavior between cities, which implies that the OPS model is more suitable for measuring the intercity interaction intensity. However, applications of the OPS model to measure intercity interactions is still lacking.
	
	In this paper, we establish an intercity interaction measurement model named the interactive city choice (ICC) model by improving the OPS model. We apply this model to measure the intercity interaction intensity in China and analyze the impact of the change in  China's land transportation network from 2005 to 2018 on the city interaction intensity. Compared with the earlier gravity model and radiation model, the results obtained by our model are more consistent with reality, implying that it captures the underlying mechanism of individual behavior in choosing to interact with cities.
	
	\section*{Interactive city choice (ICC) model}
	The OPS model\cite{LiuYan-519} assumes that when an individual chooses a destination, similar to the classic radiation model\cite{SiminiGonzALez-509} and the population-weighted opportunities model\cite{YanZhao-512,YanWang-515}, he/she first evaluates the benefit of the opportunities in each location, in which the number of opportunities in a location is proportional to the location's population, and the benefit of opportunities is a random variable with a distribution of $p(z)$ ($p(z)$ can be any continuous distribution). After evaluating the benefit, the individual will select a location that presents higher benefits than the origin and any intervening opportunities. According to the above assumption, when an individual at location $i$ makes a choice for location $j$, the probability of location $j$ being selected is
	\begin{equation}
		Q_{ij}=\int_0^{\infty}{\mathrm{Pr}_{m_{i}+s_{ij}}(z)}\mathrm{Pr}_{m_{j}}(>z) \mathrm{d}z,
	\end{equation}
	where $m_i$ is the number of opportunities at location $i$, and $s_{ij}$ is the number of intervening opportunities (i.e., the sum of the number of opportunities at all locations at a shorter distance to $i$ than $j$\cite{Stouffer-498}; see Fig. 1(a)). ${\mathrm{Pr}_{m_{i}+s_{ij}}(z)}$  is the probability that the maximum benefit obtained after $m_i+s_{ij}$ sampling is exactly $z$, and ${\mathrm{Pr}_{m_j} (>z)}$ is the probability that the maximum benefit obtained after $m_j$ samplings is greater than z.
	
	Since $\mathrm{Pr}_{x}(<z)=p(<z)^{x}$, we can obtain 
	\begin{equation}
		\mathrm{Pr}_{x}(z)=\frac{\mathrm{d}\mathrm{Pr}_{x}(<z)}{\mathrm{d}z}=x p(<z)^{x-1}\frac{\mathrm{d}p(<z)}{\mathrm{d}z}.
	\end{equation}
	Substituting Eq. (2) in Eq.(1), we obtain
	\begin{equation}
		\label{eq-3}
		\begin{aligned}
			Q_{ij}&=\int_0^{\infty}(m_i+s_{ij})p(<z)^{m_i+s_{ij}-1}\frac{\mathrm{d}p(<z)} {\mathrm{d}z} [1-p(<z)^{m_{j}}]\mathrm{d}z\\
			&=(m_i+s_{ij})\int_0^{1}{[p(<z)^{s_{ij}+m_{i}-1}-p(<z)^{m_{j}+s_{ij}+m_{i}-1}]}\mathrm{d}p(<z)\\
			&=(m_i+s_{ij}) (\frac{p(<z)^{s_{ij}+m_{i}}}{s_{ij}+m_{i}}\Big\vert_0^{1}- \frac{p(<z)^{m_j+s_{ij}+m_{i}}}{m_j+s_{ij}+m_{i}}\Big\vert_0^{1})\\
			&=(m_i+s_{ij}) (\frac{1}{s_{ij}+m_{i}}- \frac{1}{m_j+s_{ij}+m_{i}})\\
			&=\frac{m_j}{m_i+s_{ij}+m_j}.
		\end{aligned}
	\end{equation}

	\begin{figure}
		\includegraphics[width=1.0\columnwidth]{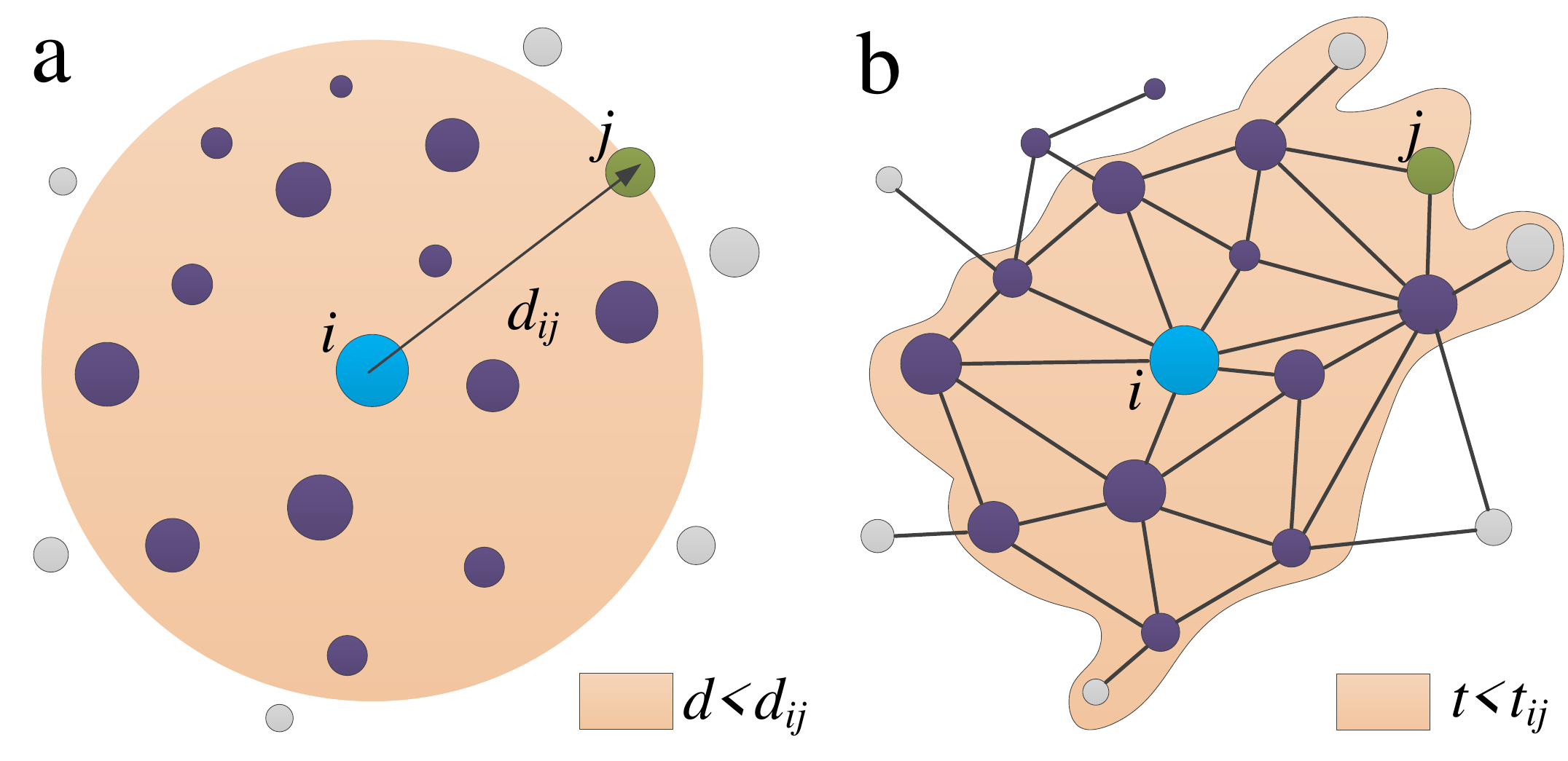}
		\caption{Schematic diagram of the calculation of intervening opportunities. Each dot represents a location. Location $i$ (blue dot) is the origin, and location $j$ (green dot) is the destination. (a) In the OPS model, the intervening opportunity $s_{ij}$ refers to the sum of the number of opportunities at all locations (i.e., locations in the orange circle, except location $i$ and location $j$) whose distance $d$ from location $i$ is less than the distance $d_{ij}$ from location $i$ to location $j$. (b) In the ICC model, intervening opportunity$s_{ij}$ refers to the sum of the number of opportunities at all locations (i.e., locations in the orange area, except location $i$ and location $j$) whose travel time $t$ from location $i$ is less than the travel time $t_{ij}$ from location $i$ to location $j$, where the travel time is obtained by calculating the shortest travel time path through the transportation network.}
	\end{figure}

	From Eq. (3), we can see that the OPS model can calculate the probability of an individual choosing a destination without any adjustable parameters. However, there are two problems in applying the OPS model to measure intercity interactions. One is that the intervening opportunities $s_{ij}$ in the OPS model are calculated by the geographic distance between two locations, as shown in Fig. 1(a). However, in reality, locations are connected by a transportation network. Most individuals compare which locations are easier to reach by taking travel time, instead of geographic distance, as the main factor. Therefore, the intervening opportunities $s_{ij}$ should be calculated by the travel time between two locations\cite{RenErcsey-Ravasz-516}, as shown in Fig. 1 (b). The other is that the OPS model assumes that the number of location opportunities is proportional to its population. However, the number of opportunities provided by each city is not directly proportional to the population but is more related to the city's industrial scale, GDP or other economic indicators\cite{ReillyOthers-568}, among which the most commonly used indicator is GDP\cite{Fagiolo-567}. Therefore, it is more reasonable to use GDP to reflect the number of opportunities. To solve these two problems, we establish a 
	probability model for individuals to choose to interact with a city. We assume that the intervening opportunities $s_{ij}$ are calculated by the travel time and that the number of location opportunities is proportional to the city's GDP. Given the GDP of each city, we can calculate the probability that an individual located in city $i$ selects city $j$ as
	\begin{equation}
		Q_{ij}=\frac{m_j}{m_i+s_{ij}+m_j},
	\end{equation}
	where $m_i$ is the GDP of city $i$, and  $s_{ij}$ is the sum of the GDP of all cities whose travel time from city $i$ is shorter than that of city $j$ (see the orange area in Fig. 1 (b)). Furthermore, if we know the total population $n_i$ of city $i$, we can calculate the interaction intensity from city $i$ to city $j$ as
	\begin{equation}
		T_{ij}={n_i}{Q_{ij}}=\frac{{n_i}{m_j}}{m_i+s_{ij}+m_j}.
	\end{equation}
	We name Eq. (5) the interactive city choice (ICC) model. It should be noted that the spatial interaction intensity $T_{ij}$ is not an actual flow volume but a dimensionless value. Furthermore, according to the spatial interaction intensity $T_{ij}$, we can calculate the interaction intensity of the city as
	\begin{equation}
		\begin{aligned}
			T_{i}=\frac{\sum\limits_{{j}\neq{i}}T_{ij}+\sum\limits_{{k}\neq{i}}T_{ki} }2,
		\end{aligned}
	\end{equation}
	where $\sum\limits_{{j}\neq{i}}T_{ij}$  is the sum of the outgoing interaction intensity and 
	$\sum\limits_{{k}\neq{i}}T_{ki}$ is that of the incoming interaction intensity\cite{SimYaliraki-513}.

	\section*{Application of the ICC model to measuring intercity interaction intensity}
	In this section, we apply the ICC model to measure the interaction intensity between cities in China and analyze the impact of the change in the Chinese land transportation network from 2005 to 2018 on city interaction intensity. It should be noted that China started to build high-speed railways in 2005; thus, we select 2005 as the starting year. Because we can only download Chinese economic and demographic data up to 2018, when we started this work, we select 2018 as the ending year.
	\subsection*{Data and processing methods}
	We select 339 Chinese cities, including 333 prefecture-level cities, four municipalities (Beijing, Tianjin, Shanghai, Chongqing) and two special administrative regions (Hong Kong and Macao), as the research objects. We download the population and GDP data of the 339 Chinese cities in 2005 and 2018 from the official website of the National Bureau of Statistics of China and the data of the cities’ central points, Chinese road networks and railway networks in 2018 from the OpenStreetMap website. The reason for selecting the road and railway network data is that the total annual transportation volume of these two land transportation modes accounts for more than 84\% of the total annual transportation volume of all intercity transportation modes (including railway, road, waterway and airway) in both 2005 and 2018, as shown in Table 1. In these two types of data, roads include highways, national roads, provincial roads, county roads and township roads; railways include high-speed railways and normal railways. Since travel between any two cities can be realized through national roads, we select the national road data as the basic land transportation network data. Furthermore, we add three other types of data (i.e., highways, normal railways and high-speed railways) that are designed to provide faster travel than national roads in the land transportation network. We establish the 2018 land transportation network, in which the edges represent highways, national roads, normal railways or high-speed railways. We add the city central point to the land transportation network by connecting it to the nearest road within the urban area. We also connect it to the nearest railway if there is a railway station within the urban area. We also need to assign the travel time value to each edge to calculate the intercity travel time in the land transportation network. We know the length of each edge in the land transportation network, so we only need to set the speeds of these four transportation modes (i.e., national road, highway, normal railway and high-speed railway) to calculate the travel time. According to the standards, including {\it Code for Design of Railway Line} (TB 10098-2017) and {\it Design Specification for Highway Alignment} (JTG D20-2017), the design speed range of high-speed railway is from 250 km/h to 350 km/h, of the normal railway is from 80 km/h to 200 km/h, of the highway is from 80 km/h to 120 km/h and of the national road is from 60 km/h to 100 km/h. For simplicity, we use the median value of the speed range, i.e., 300 km/h as the assumed speed for the high-speed railway, 140 km/h for the normal railway, 100 km/h for the highway and 80 km/h for the national road. We then calculate the travel time of each edge by dividing its length by its assumed speed. We obtain the 2005 land transportation network by deleting roads and railways built after 2005 in the 2018 land transportation network according to the 2005 Chinese road map and 2005 Chinese railway map, as shown in Fig. 2.

	\begin{table*}
		\caption{The proportion of annual 
			passengers and freight transportation 
			volume for various transportation modes 
			in China}
		\renewcommand{\arraystretch}{1.5}
		\begin{tabular}{p{3cm}p{3.6 cm}p{3.6 cm}p
				{3.6cm}p{3.6 cm}}
			\hline
			Transportation mode & 2005 Passenger transportation volume(\%) & 
			2005 freight   transportation volume(\
			\%) & 2018 Passenger   transportation 
			volume(\%) & 2018 freight 
			transportation volume(\%) \\ \hline
			railway             & 6.26              
			& 14.7         
			& 18.81     
			& 7.83 
			\\
			road                & 91.9              
			& 72.35        
			& 76.17     
			& 
			76.73                                  
			\\
			waterway            & 1.1               
			& 11.49        
			& 1.56      
			& 
			13.58                                  
			\\
			airway              & 0.75              
			& 0.02         
			& 3.4       
			& 0.01 
			\\  
			\hline    
		\end{tabular}
	\end{table*}
	
			\begin{figure}
				\includegraphics[width=1.0\columnwidth]{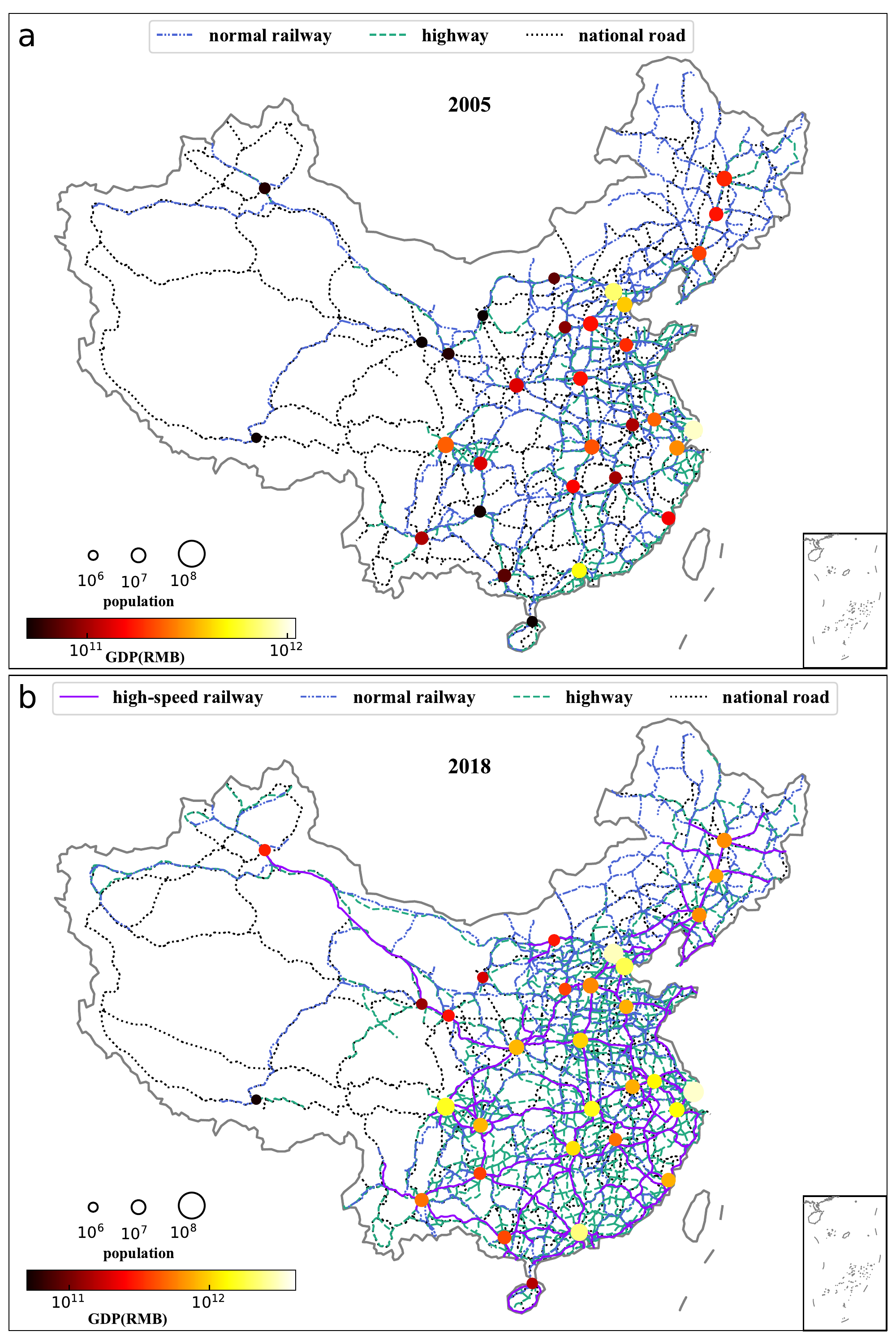}
				\caption{Chinese land transportation network and selected important cities. (a) 2005. (b) 2018. Each circle represents a provincial capital city or municipality. The size and color of the circle represent the population and GDP of the city, respectively. Solid lines represent high-speed railways, dashed-dotted lines represent normal railways, dashed lines represent highways and dotted lines represent national roads.}
			\end{figure}
			
	\subsection*{Calculation of the intercity interaction intensity}
				\begin{figure}
					\includegraphics[width=1.0\columnwidth]{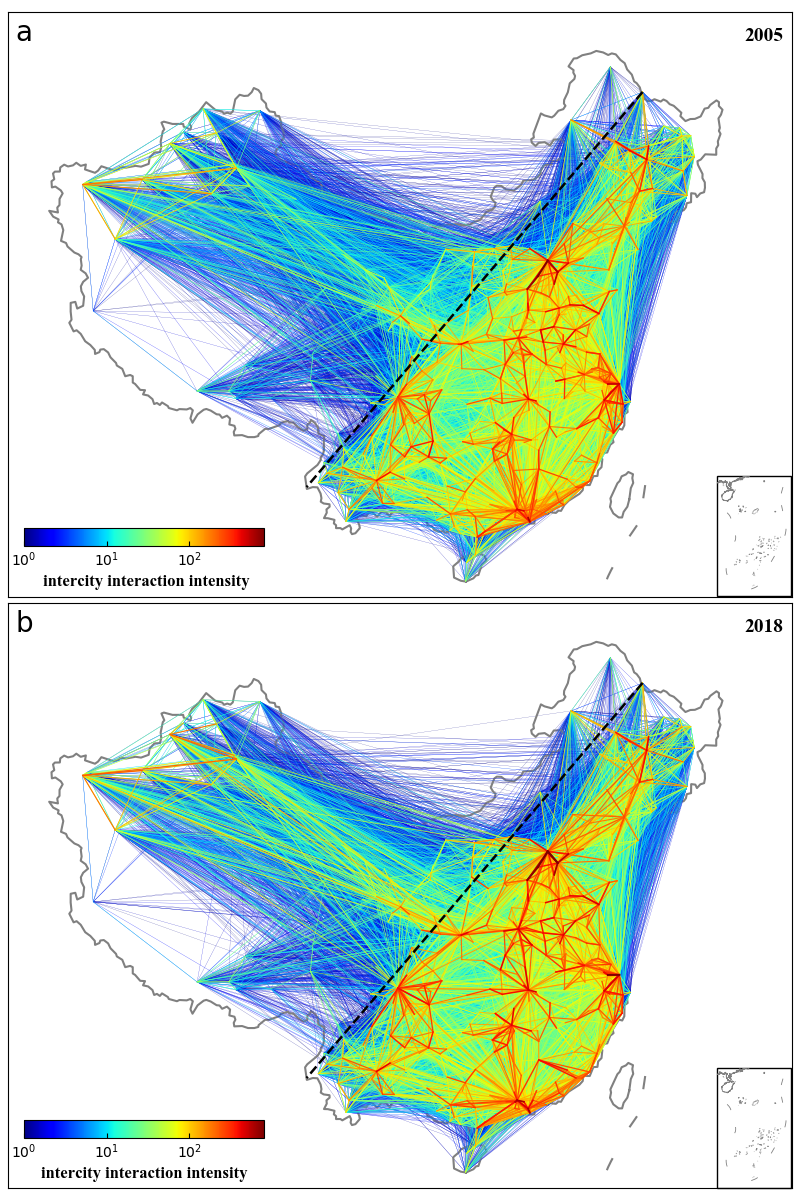}
					\caption{Distribution of interaction intensity among Chinese cities. (a) 2005. (b) 2018. The black dashed line is the Heihe-Tengchong line. The thickness and color of the other lines indicate the intercity interaction intensity.}
				\end{figure}
				
	We apply the ICC model to calculate the intercity interaction intensity in 2005 and 2018. We first calculate the travel time between cities by finding the shortest intercity travel time path in both the 2005 and 2018 land transportation networks. According to the intercity travel time, we can obtain $s_{ij}$ by summing the GDP of all cities whose travel time from city $i$ is less than the travel time from city $i$ to city $j$. We then calculate the intercity interaction intensity in 2005 and 2018 according to Eq. (5). The results are shown in Fig. 3, from which we can see that the intercity interaction intensity in the east of the Heihe-Tengchong line\cite{Huanyong-586} is higher than that in the west in both 2005 and 2018. Considering Fig. 2 and Fig. 3 comprehensively, we can see that the interaction intensity between large cities in 2018 is significantly higher than that in 2005. This increase is due to the more intensive construction of high-speed railways and highways between large cities during these 13 years. Additionally, the intercity travel time has been greatly shortened with the development of the land transportation network. We can see from Fig. 4 that the proportion of short-time interactions increases and the proportion of long-time interactions decreases from 2005 to 2018. The longest travel time was shortened from 54 h in 2005 to 43 h in 2018. Furthermore, as shown in the subgraph of Fig. 4, it only takes 7.5 h to fall within 95\% of the total national interaction intensity in the land transportation network in 2018, while it took 13 h to reach that in 2005.

		\begin{figure}
			\includegraphics[width=1.0\columnwidth]{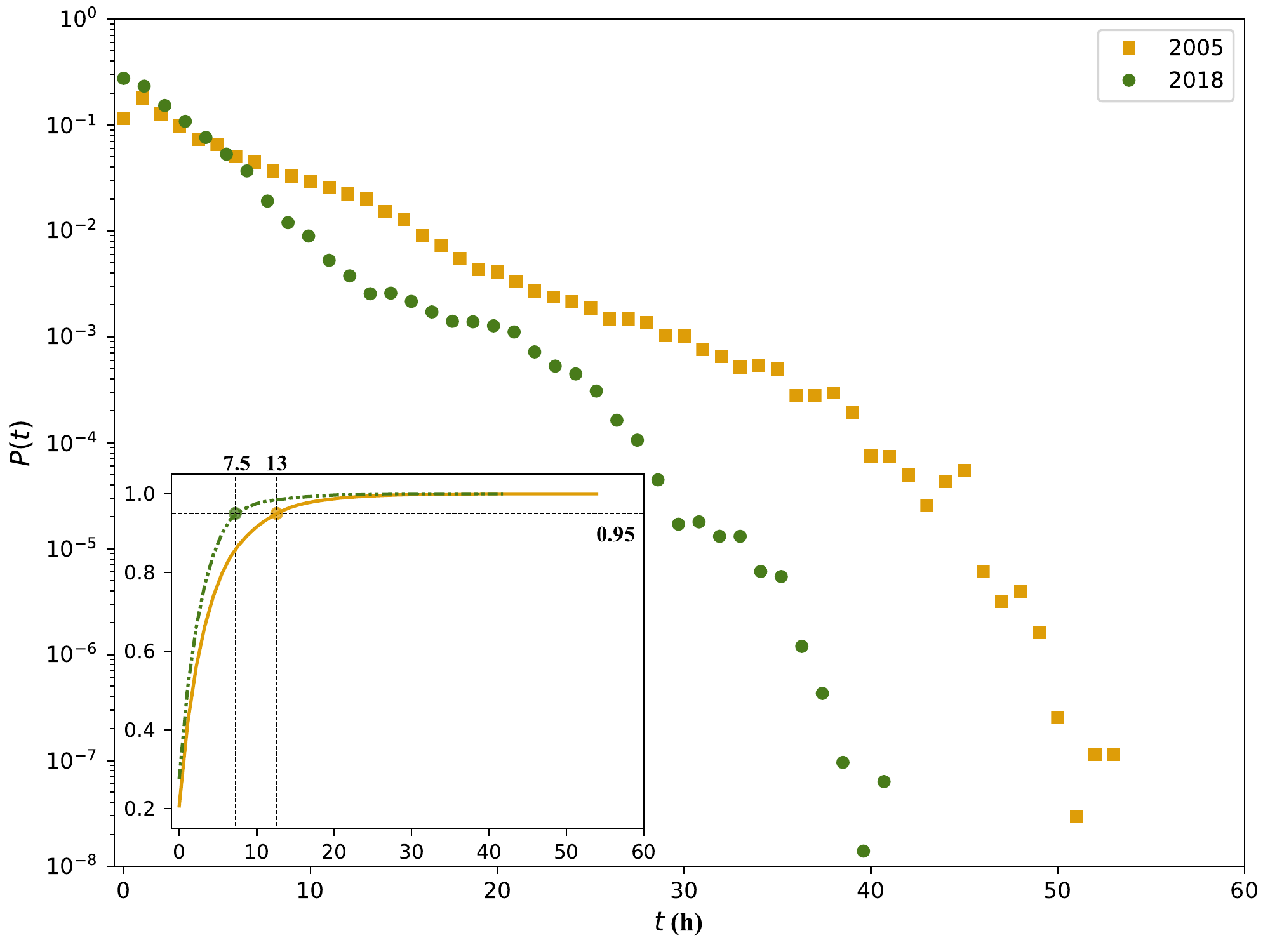}
			\caption{Travel time distribution. The square dot and circular dot in the figure represent the proportion of the intercity interaction intensity with a travel time of $t$ hours in 2005 and 2018, respectively. The solid line and dotted lines in the subgraph represent the cumulative probability distribution of travel time in 2005 and 2018, respectively.}
		\end{figure}
				
	\subsection*{Calculation of the city interaction intensity}

	\begin{figure}
		\includegraphics[width=1.0\columnwidth]{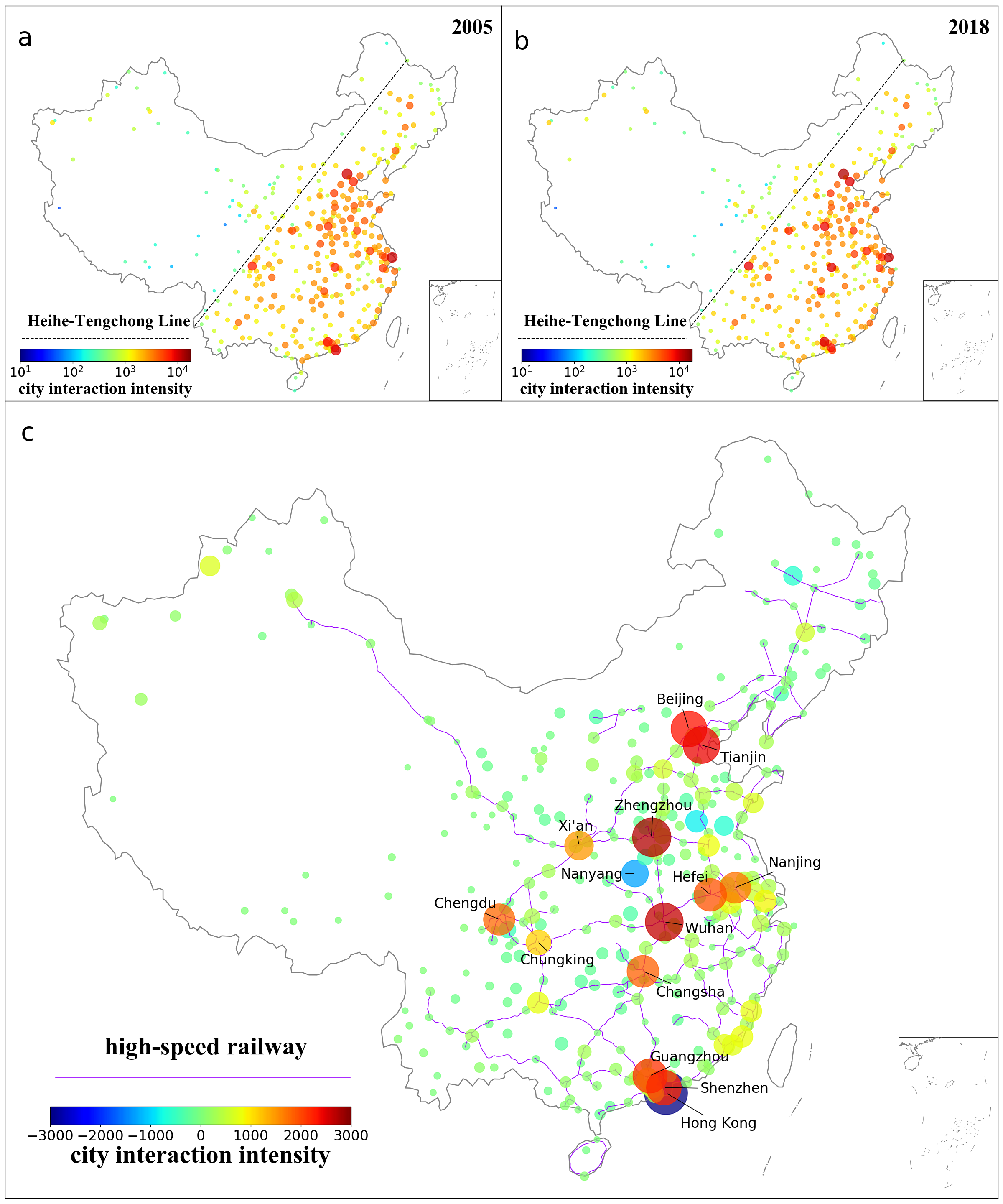}
		\caption{City interaction intensity and the difference between city interaction intensity in 2005 and that in 2018. (a) 2005 and (b) 2018 Chinese city interaction intensity. The dotted line is the Heihe-Tengchong line and each circle represents a city. The color and size of the circle indicate the city interaction intensity. (c) The difference between city interaction intensity in 2005 and that in 2018. Each circle represents a city and the size of the circle represents the absolute value of the difference value.}
	\end{figure}

	We calculate the interaction intensity of Chinese cities in 2005 and 2018 according to Eq. (6). The results are shown in Fig. 5(a-b), from which we can see that the two years show a similar distribution of city interaction intensity, i.e., the cities with high interaction intensity are mainly concentrated east of the Chinese Heihe-Tengchong Line. This is mainly because these cities have a more developed economy, more opportunities and a more intensive surrounding land transportation network, as shown in Fig. 2. Furthermore, we calculate the difference in city interaction intensity between 2005 and 2018, as shown in Fig. 5 (c). We can see that the interaction intensity of many cities, e.g., Wuhan, significantly improves, while the interaction intensity of some cities, e.g., Hong Kong, decreases. The reasons for this phenomenon are the changes in these cities’ GDP and the development of land transportation networks (especially high-speed railway networks) from 2005 to 2018. For example, Wuhan's GDP increased from 223.823 billion yuan in 2005 to 1484.729 billion yuan in 2018, with a growth rate of 663.35\%. In addition, the two national high-speed railway arteries (i.e., Beijing-Guangzhou and Shanghai-Hanrong high-speed railways) established after 2005 both pass through Wuhan. The development of Wuhan's GDP and transportation infrastructure have led to a rapid increase in its attractiveness for interaction, so the interaction intensity of Wuhan significantly improves. In contrast, Hong Kong's GDP increased from 1384.5 billion yuan in 2005 to 2400.098 billion yuan in 2018, with a growth rate of 73.35\%. It has the lowest GDP growth rate compared with other cities, which reduces its attraction for interaction. In addition, we can see from Fig. 5(c) that the interaction intensity of most cities along the high-speed railway increases, while that of cities far from the high-speed railway generally decreases. This is mainly because the travel time between cities along the high-speed railway and other cities has been significantly reduced with the construction and rapid development of the high-speed railway. Cities along the high-speed railway will be chosen with higher probability by those choosing to interact with a city.
	
		\begin{figure}
			\includegraphics[width=1.0\columnwidth]{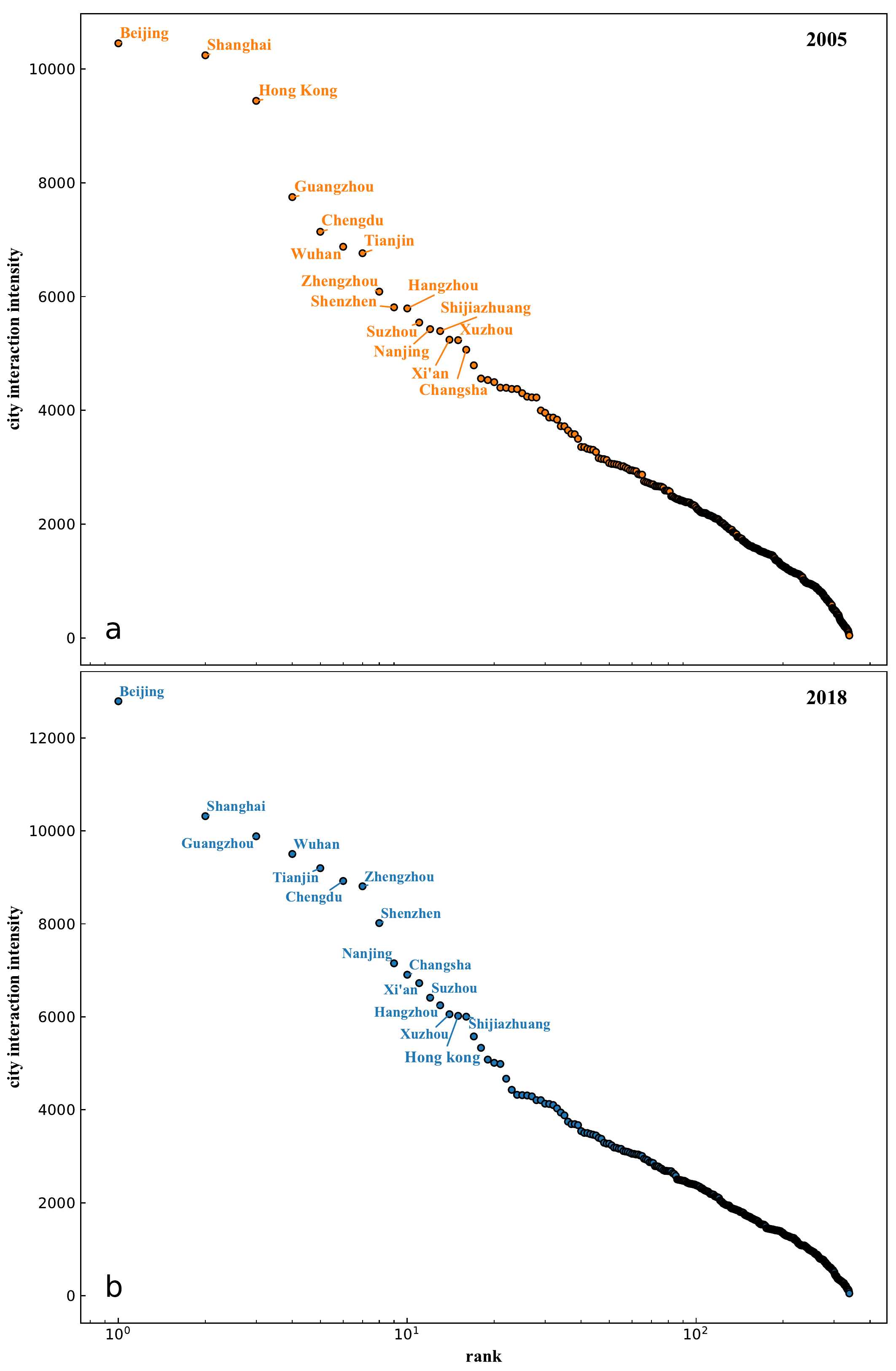}
			\caption{Ranking of city interaction intensity. (a) 2005. (b) 2018. The horizontal axis represents the ranking of cities, and the vertical axis represents the city interaction intensity.}
		\end{figure}	
		\begin{figure}
			\includegraphics[width=1.0\columnwidth]{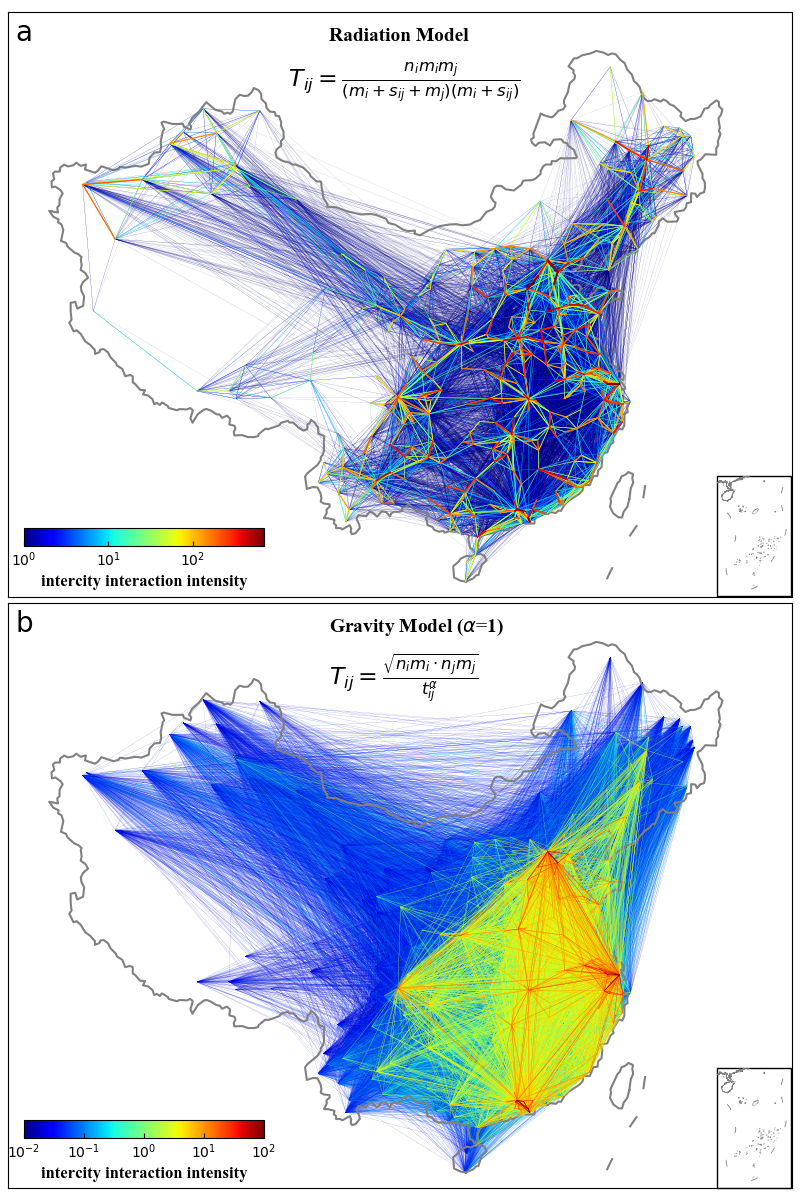}
			\caption{The distribution of the intercity interaction intensity calculated by different models. (a) Radiation model. (b) Gravity model. The thickness and color of the lines indicate the intercity interaction intensity.}
		\end{figure}

	We further rank the 339 cities according to their interaction intensity in 2005 and 2018. The results are shown in Fig. 6, from which we can see that the ranking of city interaction intensity changes greatly from 2005 to 2018. For example, Guangzhou's ranking rises from fourth place to third place, Shenzhen's ranking rises from ninth place to seventh place, and Hong Kong's ranking drops from third place to fifteenth place. Although these three cities are geographically close and their 2018 GDPs are similar (as shown in Table 2), they differ greatly in their city interaction intensity due to their different positions in the land transportation network. Guangzhou is one of three national comprehensive transportation hubs with three national road arteries (i.e., Beijing-Guangzhou Line, Guiyang-Guangzhou Line and Nan-Guangzhou Line). Shenzhen is also a comprehensive transportation hub that connects Hong Kong, Macao and mainland China.

\begin{table*}
	\caption{Ranking of population, 
		GDP and city interaction intensity 
		calculated by the ICC model, radiation 
		model and gravity model}
	\renewcommand{\arraystretch}{1.5}
	\begin{tabular}{p{1.4cm}p
			{2.4cm}p{2.4cm}p{2.4cm}p{2.4cm}p
			{2.4cm}p{2.4cm}}
		\hline
		Rank&Population   & GDP 
		& ICC model    & Radiation model 
		& Gravity model($\alpha$=1) & Gravity   
		model($\alpha$=2) \\ \hline
		1    & Shanghai     & 
		Shanghai  & Beijing      & Beijing      
		& Shanghai           & Shenzhen      
		\\
		2    & Beijing      & 
		Beijing   & Shanghai     & Guangzhou    
		& Beijing            & Hong Kong     
		\\
		3    & Chengdu      & 
		Shenzhen  & Guangzhou    & Zhengzhou    
		& Shenzhen           & Guangzhou     
		\\
		4    & Tianjin      & 
		Hong Kong & Wuhan        & Tianjin      
		& Guangzhou          & Foshan        
		\\
		5    & Guangzhou    & 
		Guangzhou & Tianjin      & Chengdu      
		& Tianjin            & Shanghai      
		\\
		6    & Shenzhen     & 
		Tianjin   & Chengdu      & Shanghai     
		& Hong Kong          & Beijing       
		\\
		7    & Baoding      & 
		Suzhou    & Zhengzhou    & Wuhan        
		& Suzhou             & Hangzhou      
		\\
		8    & Wuhan        & 
		Chengdu   & Shenzhen     & Hangzhou     
		& Wuhan              & Tianjin       
		\\
		9    & Shijiazhuang & 
		Wuhan     & Nanjing      & Xi'an        
		& Hangzhou           & Suzhou        
		\\
		10   & Suzhou       & 
		Hangzhou  & Changsha     & Shenzhen     
		& Nanjing            & Nanjing       
		\\
		11   & Linyi        & 
		Nanjing   & Xi'an        & Changsha     
		& Zhengzhou          & Xi'an         
		\\
		12   & Zhengzhou    & 
		Qingdao   & Suzhou       & Shijiazhuang 
		& Chengdu            & Zhengzhou     
		\\
		13   & Nanyang      & 
		Wuxi      & Hangzhou     & Nanjing      
		& Foshan             & Dongguan      
		\\
		14   & Xi'an        & 
		Changsha  & Xuzhou       & Xuzhou       
		& Changsha           & Wuhan         
		\\
		15   & Hangzhou     & 
		Ningbo    & Hong Kong    & Shenyang     
		& Wuxi               & Shaoxing      
		\\
		16   & Handan       & 
		Zhengzhou & Shijiazhuang & Suzhou       
		& Hefei              & Wuxi          
		\\
		17   & Harbin       & 
		Foshan    & Hefei        & Harbin       
		& Xuzhou             & Xianyang      
		\\
		18   & Qingdao      & 
		Quanzhou  & Chungking    & Jinan        
		& Ningbo             & Chengdu       
		\\
		19   & Weifang      & 
		Nantong   & Qingdao      & Changchun    
		& Shijiazhuang       & Jiaxing       
		\\
		20   & Wenzhou      & 
		Chungking & Foshan       & Handan       
		& Dongguan           & Changsha  \\ 
		\hline    
	\end{tabular}
\end{table*}

	We next apply the radiation model\cite{SiminiGonzALez-509} and gravity model\cite{JiaoTang-580} to calculate the intercity interaction intensity (see Fig. 7) and the interaction intensity of each city in 2018. We further list the ranking of the interaction intensity of the top 20 Chinese cities calculated by the ICC model, radiation model and gravity model (see Table 2). Compared with the radiation model and gravity model, the result of the ICC model is more consistent with the actual situation. This is because the radiation model assumes that when the individual chooses an interactive city, he/she only chooses the nearest city with a higher GDP than his/her home\cite{SiminiGonzALez-509}; the gravity model assumes that the intercity interaction intensity is proportional to the product of the two cities’ size (population or GDP)\cite{JiaoTang-580} and inversely proportional to a power function of the travel time with a power exponent of 1 or 2\cite{Odlyzko-581}. Neither the radiation model nor the gravity model can accurately capture the mechanism of individuals choosing to interact with cities, so they cannot reasonably reflect the city interaction intensity. Unlike the radiation model and gravity model, the ICC model's ranking of the cities’ interactive intensity suggests that our model can better describe the essential mechanism behind the individual choice to interact with a city.

	\section*{Conclusion and discussion}
	The measurement of intercity interaction has important research significance. In this paper, we develop the ICC model from the perspective of individual choice behavior, which assumes that the probability of an individual choosing to interact with a city is proportional to the number of opportunities, as expressed by the GDP of the destination city, and inversely proportional to the number of intervening opportunities, calculated by the shortest travel time in the land transportation network. Multiplying this probability by the origin city's population, one can obtain the intercity interaction intensity. To demonstrate the advantage of the ICC model, we apply the ICC model to measure the interaction intensity among 339 cities in China. After collecting and processing the big data related to intercity interaction, we analyze the impact of the change in the land transportation network from 2005 to 2018 on the intercity and city interaction intensity. We find that the travel time between cities has decreased and the interaction intensity between large cities has increased due to the development of land transportation. In particular, the interaction intensity of cities along high-speed railways has greatly increased. Compared with the previously widely used spatial interaction models (i.e., the gravity model and radiation model), the results obtained by the ICC model are more consistent with the actual situation.
	
	The proposed ICC model not only helps us better measure the intercity interaction intensity but also offers potential additional applications. For example, the ICC model provides a new perspective for identifying suburbs, which is a hot topic in geographical research. The traditional suburban identification method usually refers to population density and the nature of residential land\cite{Tian-562}. The ICC model introduces the spatial interaction intensity between city districts, which can improve the method of suburban identification. In addition, the ICC model can calculate the interaction intensity within and between urban agglomerations, providing valuable indicators for a comprehensive evaluation of the degree of urban agglomeration\cite{LiFeng-561,ShiYan-572}, which is of great significance for urban agglomeration sustainable development.
	
	Although the ICC model can obtain more realistic results when measuring intercity interaction intensity, it still has room for expansion in practical applications. In this paper, we use GDP, which is a key factor affecting the number of opportunities, to reflect the number of opportunities. In reality, there are many other factors, e.g., urban population, industrial size and industrial structure, that also affect a city's opportunities. Therefore, we can use multiple factors to calculate the number of opportunities in future applications. In addition, we only use the travel time calculated by the shortest time path algorithm in the land transportation networks, including roads and railways, to measure the interaction intensity among 339 Chinese cities. However, the importance of various transportation modes is different in different countries or regions. For example, airways are an important mode of passenger transportation between the U.S.  cities\cite{zhang2017energy}, and waterways are the main mode of freight transport between European cities\cite{benga2019assesment}. Therefore, future research can consider extending the land transportation network to a more comprehensive three-dimensional transportation network including roads, railways, airways and waterways to make more reasonable measurements of intercity interactions.
	
	\section*{Acknowledgements}
	We thank Professor Kai Liu and her postgraduate student Wei-Hua Zhu at Beijing Normal University for their help with our early research.
	
	\section*{Funding}
	This work was supported by the National Natural Science Foundation of China (Grant Nos. 71822102, 71621001, 71871010). 
	
\bibliographystyle{bmc-mathphys}
\bibliography{bmc_article}

\end{document}